\documentclass[preprint,12pt]{elsarticle}

 \usepackage{graphics}
\usepackage{graphicx}
\usepackage{amssymb}
\usepackage{amsthm}
\usepackage[T1]{fontenc}
\usepackage{geometry}
\usepackage{amsmath}
\usepackage{mathrsfs}
\usepackage[all]{xy}
\usepackage{algorithm} 
\usepackage{algorithmic} 
\usepackage{url}
\usepackage{color}
\usepackage{appendix}
\usepackage{algorithm} 
\usepackage{algorithmic} 
\usepackage{url}
\usepackage{pxfonts}
\usepackage{dsfont}

\newtheorem{Theo}{Theorem}%[section]
\newtheorem{Prop}{Proposition}
\newtheorem{Proof}{Proof}
\newtheorem{Rem}{Remark}
\newtheorem{Def}{Definition}

\RequirePackage{cite}
\RequirePackage[numbers]{natbib}
\RequirePackage[colorlinks,citecolor=blue,urlcolor=blue]{hyperref}

\journal{Insurance: Mathematics and Economics}

\begin{document}

\begin{frontmatter}


\title{Explicit Expressions for Multidimensional Value-at-Risk under Archimedean Copulas}


%\author{Amadou Sawadogo}

%\address{}
\author[label1=e1]{Dotamana \textsc{Yéo}} %%\ead{dotamanayeo8@gmail.com}
\author[label2=e2]{Saralees  \textsc{Nadarajah}}%%\ead{Saralees.Nadarajah@manchester.ac.uk}
\author[label1=e1]{Amadou \textsc{Sawadogo}}
\ead{amadou.sawadogo@gmail.com}
%% \ead{reolie.mizele@udsn.cg}%\footnote{Corresponding author}
\address[label1=e1]{Universit\'e F\'elix Houphou\"et Boigny,  UFR de Math\'ematiques et Informatique, 22 BP 582 Abidjan 22, C\^ote d'Ivoire.}
\address[label2=e2]{University of Manchester, Manchester M13 9PL, UK.}
% \address[label1=e1]{Universit\'e F\'elix Houphou\"et Boigny,  UFR de Math\'ematiques et Informatique, % 22 BP 582 Abidjan 22, C\^ote d'Ivoire.}

%\ead{ramananantoandro@gmail.com}

%\cortext[mycorrespondingauthor]{Corresponding author at: UFR de Math\'ematiques et Informatique, Universit\'e  F\'elix Houphou\"et Boigny, 22 BP 582 Abidjan 22, C\^ote d'Ivoire}
%\ead{amadou.sawadogo@gmail.com}
%\author[cck]{C\'elestin C. \textsc{Kokonendji}}
%\address[cck]{Laboratoire de math\'ematiques de Besan\c con, Universit\'e Bourgogne Franche-Comt\'e, Besan\c con, France}
%\ead{celestin.kokonendji@univ-fcomte.fr}

\begin{abstract}
This paper studies multivariate Value-at-Risk (VaR) for financial portfolios with a focus on modeling dependence structures through Archimedean copulas. Using the generator representation of Archimedean copulas, we derive explicit analytical expressions for the marginal lower-tail multivariate VaR in arbitrary dimensions.

Closed-form formulas are obtained for several commonly used copula families, including Clayton, Frank, Gumbel-Hougaard, Joe and Ali--Mikhail--Haq copulas, allowing a direct assessment of the impact of dependence on multivariate risk. These results complement existing approaches, which largely rely on numerical or simulation-based methods, by providing tractable alternatives for theoretical and applied risk analysis.

Monte Carlo simulations are conducted to evaluate the finite-sample performance of the proposed VaR estimator and to illustrate the role of different dependence structures. The proposed analytical setting  offers transparent  tools for multivariate risk measurement and systemic risk assessment.
\end{abstract}


\end{frontmatter}

% \linenumbers

% \linenumbers


\section{Introduction}

The accurate measurement of financial risk in modern portfolios requires risk measures capable of accounting for complex dependence structures among multiple risk components. While univariate risk measures such as the Value-at-Risk (VaR) remain central tools in regulatory and practical risk management, they fail to capture the joint behavior of portfolio constituents and may therefore underestimate aggregate risk in the presence of dependence. This limitation has motivated an extensive body of research devoted to the development of multivariate and vector-valued risk measures that incorporate dependence effects in a coherent and transparent manner.

Among the available approaches, copula-based models have emerged as a powerful and flexible approach for describing multivariate dependence independently of marginal distributions (Hofert et al., 2018). Since the seminal work of Sklar (1959), copulas have been widely used in finance and insurance to model joint distributions with heterogeneous marginals and complex dependence patterns. In particular, Archimedean copulas have attracted considerable attention due to their analytical tractability, ease of implementation, and ability to model a wide range of dependence structures through a single generator function. Comprehensive treatments of copula theory can be found in Nelsen (2006) and Joe (1997), among others.

Parallel to the development of copula theory, the literature on multivariate risk measures has grown significantly over the past two decades. Several extensions of univariate VaR and Conditional Tail Expectation (CTE) to the multivariate setting have been proposed, including vector-valued risk measures, set-valued approaches, and aggregation-based methods. Notable contributions include H{\"u}rlimann (2017) and subsequent works, which investigate multivariate VaR and CTE within copula frameworks and emphasize the role of dependence in determining joint tail risk. Despite these advances, many existing results rely heavily on numerical integration or Monte Carlo simulation, particularly when dealing with higher-dimensional portfolios or non-Gaussian dependence structures.

In this context, analytical expressions for multivariate risk measures remain relatively scarce. Closed-form or semi-analytical representations are typically available only under restrictive assumptions or for specific dependence models. This lack of tractability limits both theoretical insight and practical applicability, as simulation-based approaches may be computationally intensive and provide limited transparency regarding the impact of dependence parameters on the resulting risk measures. Recent contributions include, among others, Guégan and Hassani (2019), Hoffmann et al. (2018), Lux and Papapantoleon (2019), and  Qian et al. (2019), who investigate multivariate risk measures and dependence effects under various modeling frameworks.

This paper addresses these limitations by focusing on multivariate Value-at-Risk within the class of Archimedean copulas. The generator representation of Archimedean copulas together with Kendall’s distribution, which naturally characterizes their level sets and plays a central role in the definition of multivariate quantiles are exploited. We derive explicit analytical expressions for the marginal lower-tail multivariate VaR in arbitrary dimensions.

Closed-form formulas are obtained for several widely used Archimedean copula families, including the Clayton, Frank, Gumbel-Hougaard, Joe and Ali-Mikhail-Haq copulas. These results provide a clear and direct link between the copula dependence parameter and the resulting multivariate risk measure, thereby offering analytical insight into the effect of dependence on portfolio risk. In contrast to approaches that rely primarily on numerical methods, the proposed derivation  yields tractable expressions that can be readily implemented and analyzed.

To assess the practical relevance of the theoretical results, we complement the analytical developments with Monte Carlo simulation experiments. These simulations evaluate the finite-sample performance of the proposed VaR estimator and illustrate how different dependence structures influence multivariate risk assessment. The numerical results confirm the accuracy of the analytical expressions and highlight the importance of properly modeling dependence in multivariate risk measurement.

The remainder of the paper is organized as follows. Section~\ref{ArchCop} reviews the necessary background on Archimedean copulas and  multivariate Value-at-Risk. Section~
\ref{ArchCp} presents the main theoretical results and derives explicit expressions for the multivariate VaR under various Archimedean copula families. Section~\ref{IllusVarApp} is devoted to Monte Carlo simulations that illustrate and validate the proposed methodology. Section~\ref{ConcRem} concludes and outlines directions for future research.

\section{Preliminaries on Archimedean copulas and multivariate Value-at-Risk}~\label{ArchCop}
This section provides an overview of selected definitions and theorem that are well established in the existing literature.
\begin{Def} (Nelsen, 2006)

A copula $C: [0,1]^{d} \longrightarrow [0,1]$ is said to be Archimedean if there exists a function $\phi: [0,1] \longrightarrow [0, \infty)$ called a generator, such that $\phi(1)=0$, $\phi$ is strictly decreasing and 
	\begin{equation} 
		C(u_{1},\cdots,u_{d})=\phi^{-1}(\phi(u_{1})+\dots+\phi(u_{d})),\;\forall (u_{1},\cdots,u_{d})\in [0,1]^{d}. 
	\end{equation} 
Moreover, in order  to be a valid copula in dimension $d$, the generator must be completely monotone on
$[0,1]$, \textit{i.e.}:
	\begin{align*}
		 (-1)^{k}\frac{d^{k}\phi^{-1}(t)}{dt^{k}}\geq 0, \, \forall \, t\in[0;\infty ) \, and \, k=0, 1,\cdots  
	\end{align*}
\end{Def}

In the following, the considered random vectors $\textbf{X}=(X_{1},\cdots,X_{d})$ meet the so-called \textit{regularity conditions}, which means that $\textbf{X}=(X_{1},\cdots,X_{d})$ is non-negative absolutely constinuous (w.r.t the Lebesgue measure on $\mathbb{R}^{d}$) with a partially increasing multivariate distribution function $F$ provided that $\mathbb{E}[X_{i}]< \infty,\;i=1,\cdots,d$.

%Hereafter,  the following definitions given in  (Di Bernardino, 2011) and  (Nikiéma, 2025) are assumed.  
\begin{Def}(Di Bernardino, 2011)

Let $\alpha\in (0,1)$ and $\mathcal{L}^{F}_{\alpha}=\{(x_{1},\cdots,x_{d})\in \mathbb{R}_{+}^{d}: F(x_{1},\dots,x_{d})\geq \alpha)\}$, the upper $\alpha$-level set of $F$. The boundary of the set  $\mathcal{L}^{F}_{\alpha}$ is  the $\alpha$-level set of $F$  expressed as  
\[ \partial\mathcal{L}^{F}_{\alpha}=\{(x_{1},\cdots,x_{d})\in \mathbb{R}_{+}^{d}: F(x_{1},\dots,x_{d})=\alpha \}. 
\] 
Using Sklar’s theorem, $\partial \mathcal{L}^{F}_{\alpha}$ admits a representation in terms of the associated copula $C$ given by
\begin{equation} 
\partial\mathcal{L}^{C}_{\alpha}=\{ (u_{1},\cdots,u_{d})\in [0,1]^{d}: C(u_{1},\cdots,u_{d})=\alpha \}. 
\end{equation}
\end{Def}

The multivariate Value-at-Risk functional to the  confidence level $\alpha\in (0,1)$ is defined as follows (Cousin and Di Bernadino, 2013, Definition 2.1, lower-orthant VaR; H{\"u}rlimann, 2017).
\begin{Def}\label{VaR:def}

Let $\textbf{X}=(X_{1},\cdots,X_{d})$ be a random vector satisfying the regularity conditions. For $\alpha\in (0,1)$, the multivariate Value-at-Risk functional to the  confidence level $\alpha$ is given by
	\begin{equation}\label{VaR:Deff}
		VaR_{\alpha}(\textbf{X})=\mathbb{E}[\textbf{X}|\textbf{X}\in\mathcal{\partial{L}}_{\alpha}^{C}]=\begin{pmatrix}
			\mathbb{E}[X_{1}|\textbf{X}\in\mathcal{\partial{L}}_{\alpha}^{C} ]	\\
			\mathbb{E}[X_{2}|\textbf{X}\in\mathcal{\partial{L}}_{\alpha}^{C} ]	\\
			\vdots\\
			\mathbb{E}[X_{d}|\textbf{X} \in\mathcal{\partial{L}}_{\alpha}^{C} ]
		\end{pmatrix}.
	\end{equation}
\end{Def}
Thus, to obtain the Value-at-Risk in a multidimensional framework, we only need to determine the marginal Value-at-Risk for all $i=1,\cdots,d$. The following theorem by  H{\"u}rlimann (2017) gives  a general  result in the case of Archimedean copulas.
\newpage
\begin{Theo}~\label{VaR:Arch}(H{\"u}rlimann, 2017)

Let $\textbf{X} = (X_1, \cdots, X_d)$ be a random vector and $F$ its  joint distribution function
associated to an absolutely continuous Archimedean  copula $C$ with generator $\phi$.
Then the components $\mathrm{VaR}_{\alpha}^{i}(\textbf{X}), i=1,\cdots,d$ of the multivariate $\mathrm{VaR}_{\alpha}(\textbf{X})$, satisfy the following formula,
\begin{equation}	
VaR_{\alpha}^{i}(\textbf{X}) = \frac{d-1}{\phi(\alpha)^{d-1}}  \int_{\alpha}^{1} VaR_{u}(X_i) \beta_{d}(u,\alpha) du,
\end{equation}
with  $\beta_{d}(u,\alpha)=-\phi^{\prime}(u).[\phi(\alpha)-\phi(u)]^{d-2},\; 1\leq u\leq 
\alpha, 0 <\alpha < 1$. 
\end{Theo}
\begin{Proof}
For the proof of theorem~\ref{VaR:Arch}, see (H{\"u}rlimann, 2017).
\end{Proof}
 The application of Theorem~\ref{VaR:Arch} enables the derivation of explicit expressions for the multivariate Value-at-Risk related to several classes of Archimedean copulas.
\section{Expression of the VaR for some Archimedean copulas}~\label{ArchCp}
This section is devoted to the analytical derivation of closed-form multivariate Value-at-Risk expressions for several widely used Archimedean copula families, namely Clayton, Frank, Gumbel-Hougaard, Joe, and Ali–Mikhail–Haq (Nelsen, 2006; Nadarajah et al., 2017).

Since the derivation of the multivariate VaR relies on the same integral representation for all Archimedean copulas, only the proof for the Clayton copula is detailed below, while the remaining proofs follow analogously.
\subsection{Clayton's family of  copulas and the VaR}
The Clayton copula is one of the most widely used Archimedean copula families in multivariate dependence modeling. Its recent applications  include (Nadarajah et al., 2017): analysis of bivariate truncated data;  tail dependence estimation in financial market risk management; probable modeling of hydrology data and estimation of failure probabilities in hazard scenarios.  It is defined as follows (Nadarajah et al., 2017; Nelsen, 2006).

\begin{Def}

For all $(u_{1}, \ldots, u_{d}) \in [0,1]^{d}$ and for any $\theta > 0$, the copula is given by  
	\[
	C_{\theta}(u_{1}, \ldots, u_{d})
	= \left ( \sum_{i=1}^{d} u_{i}^{-\theta} - d + 1 \right)^{-1/\theta}.
	\]
The associated generator is expressed as (Nelsen, 2006)
\begin{equation}
		\phi_{\theta}(t)=\frac{1}{\theta} \bigl(t^{-\theta} - 1\bigr), \; \forall t \in [0,1]
\end{equation}
with $\theta > 0$ the dependence parameter. Its tau of Kendall is given by (Nelsen, 2006) $\tau (\theta)=\theta/(\theta +2),\;\theta > 0.$
\end{Def}
\begin{Rem}

One can notice that, independence is obtained in the limit as  $\theta \to 0$, whereas complete dependence arises as $\theta \to \infty$. For $\theta >0$, the copula exhibits positive dependence with lower tail dependence.
\end{Rem}

\begin{Prop}~\label{VaRC}
Consider a random vector $\mathbf{X} = (X_{1}, \cdots, X_{d})$ with $d \geq 2$, 
whose joint distribution function $F$ satisfies the usual regularity conditions 
and is associated with a Clayton's family of copulas generated by $\phi_{\theta}(t) = \bigl(t^{-\theta} - 1\bigr)/\theta,\; \forall t \in [0,1]$ with $\theta > 0$. For any confidence level $\alpha \in (0,1)$ and dependence parameter $\theta > 0$, the marginal $\alpha$-level conditional Value-at-Risk of component $X_{i}$ is given by
\begin{equation}
\mathrm{VaR}_{\alpha}^{\,i}(\textbf{X})
=\frac{(d-1)\theta}{(\alpha^{-\theta} - 1)^{d-1}}
\int_{\alpha}^{1}
\mathrm{VaR}_{u}(X_{i}). \,
u^{-\theta - 1} \,
\bigl(\alpha^{-\theta} - u^{-\theta}\bigr)^{d-2} \,
du.
\end{equation}

\end{Prop}
\begin{Proof}
For  $\textbf{X}=(X_{1},\dots,X_{d})$ a $d$-dimensional random vector with joint   distribution  $F$ satisfying the standard regularity conditions, one has from  theorem~\ref{VaR:Arch}
	\begin{equation*}
\mathrm{VaR}_{\alpha}^{i}(\textbf{X})=\frac{d-1}{(\phi(\alpha))^{d-1}}\int_{\alpha}^{1}VaR_{u}(X_{i}). \beta(u, \alpha)du,
	\end{equation*}
with $\beta_{d}(u, \alpha)=-\phi^{\prime}(u)[\phi(\alpha)-\phi(u)]^{d-2}.$
 Using the generator of Clayton's family of copulas formula, one obtains $\beta_{d}(u, \alpha)=u^{-\theta -1}(\alpha^{-\theta}-u^{-\theta})^{d-2}/\theta^{d-2}$.
Thus,
	\begin{align*}
	\mathrm{VaR}_{\alpha}^{i}(\textbf{X})&=\frac{d-1}{(\phi(\alpha))^{d-1}}\int_{\alpha}^{1}VaR_{u}(X_{i}) \beta(u, \alpha)du\\
	&=\frac{d-1}{(\phi(\alpha))^{d-1}\theta^{d-1}}\int_{\alpha}^{1}VaR_{u}(X_{i})u^{-\theta -1}(\alpha^{-\theta}-u^{-\theta})^{d-2}du\\
	&=\frac{\theta(d-1)}{(\alpha^{-\theta}-1)^{d-1}}   \int_{\alpha}^{1}VaR_{u}(X_{i}).u^{-\theta-1}(\alpha^{-\theta}-u^{-\theta})^{d-2}du.
	\end{align*}
\end{Proof}	
For a portfolio whose dependence structure is characterized by a Clayton copula, the marginal Value-at-Risk is determined by the marginal quantile functions of the individual components. This property allows for the derivation of a closed-form expression for the marginal Value-at-Risk in the case of uniformly distributed random variables in dimension $d \geq 2$.
\begin{Prop}~\label{VaRClUn}
Let $\mathbf{X} = (X_{1}, \cdots, X_{d})$ be a random vector associated with a 
Clayton's family of copulas, where each marginal component $X_{i}$, $i = 1, \cdots, d$, follows 
a uniform distribution on $[0,1]$. For any pair $i, j \in \{1, \dots, d\}$ with 
$i \neq j$, one has $\mathrm{VaR}_{\alpha}^{\,i}(\mathbf{X})=\mathrm{VaR}_{\alpha}^{j}(\mathbf{X}),
\; \forall \alpha \in (0,1).$
Moreover, for each $i = 1, \dots, d$, the marginal  Value-at-Risk is given by
\begin{equation}~\label{Uniform:clayton}
		\mathrm{VaR}_{\alpha}^{i}(\textbf{X})
		=
		\frac{\theta \,(d-1)}{\left(\alpha^{-\theta} - 1 \right)^{d-1}}
		\int_{\alpha}^{ 1}
		\left(\alpha^{-\theta} - u^{-\theta}\right)^{d-2}
		u^{\, -\theta}\, du.
	\end{equation}
\end{Prop}

This result  is independent of the index of the random variable $X_i$. Consequently, the lower Value-at-Risk at level $\alpha$ is identical across all components, \textit{i.e.}, $\text{VaR}_{\alpha}^{i}(\textbf{X}) = \text{VaR}_{\alpha}^{j}(\textbf{X})$ for any $i, j \in \{1, \cdots, d\}$ such that 
$i \not= j$.

%Proposition~\ref{VaRClUn} extends that of Di Bernardino (2011), which gives an explicit expression of the VaR in a two-dimensional framework for a dependence structure modeled by the Clayton's copula. 

\subsection{Frank's family of copulas and the VaR}
The Frank copula has been widely employed in a broad range of applications. Recent applications include the following (Nadarajah et al., 2017): intensity-duration model of storm rainfall; analytical calculation of storm volume statistics; characterization of temporal structure of storms; modeling of higher-order correlations of neural spike counts; drought frequency analysis; modeling of acoustic signal energies. 
Frank's copula is expressed as follows (Nikiéma, 2025; Nadarajah et al., 2017).
\begin{Def}
For all  $(u_{1},\cdots,u_{d})\in[0,1]^{d}$ and $\theta \in (-\infty,\infty) \setminus  \{0 \} $, the Frank's family of copulas is defined as 
	\[ 
	C_{\theta}(u_{1},\dots,u_{d})=-\frac{1}{\theta}ln\biggl[1+\frac{\prod_{i=1}^{d}(e^{-\theta u_{i}}-1)}{(e^{-\theta}-1)^{d-1}}\biggr].
	\]
The associated generator function is as follows (Nelsen, 2006),
\begin{equation}
		\phi_{\theta}(t)=-ln\biggl(\frac{e^{-\theta t}-1}{e^{-\theta}-1}\biggr),\; \forall  t \in [0,1]
\end{equation}	 
with dependence parameter $\theta \in (-\infty,\infty) \setminus  \{0 \} $.  The corresponding tau of Kendall  is given by 
 $\tau (\theta)=1-4/\theta[1-(1/\theta)\int_{0}^{\theta}t/(e^{t}-1)dt],\;\theta \not= 0$ (Nelsen, 2006; Thilini, 2022).
\end{Def}
\begin{Rem}
The Frank copula  allows for both positive and negative dependence, governed by the parameter $\theta$
(Nadarajah et al., 2017). It is symmetric and does not exhibit tail dependence. Independence is recovered as $\theta \to 0$, while perfect dependence is obtained in the limits $\theta \to +\infty$ and $\theta \to -\infty$, respectively.
\end{Rem}
\begin{Prop}~\label{VaRFranck}
Let $\textbf{X} = (X_1, \cdots, X_d)$ be a random vector with joint distribution $F$, satisfying regularity conditions and associated with a Frank's family of copulas  characterized by the generator function 
$\phi_{\theta}(t)=-ln[(e^{-\theta t}-1)/(e^{-\theta}-1)],\;\forall t \in [0,1]\;  \mathrm{with} \; 0<\theta<\infty.$ Then, $\forall i \in \{1,\cdots,d \}$
\begin{equation}
\mathrm{VaR}^i_{\alpha}(\mathbf{X})
=
\frac{d-1}{
\biggl[-ln\biggl(\frac{e^{-\theta \alpha}-1}{e^{-\theta}-1}\biggr)\biggr]^{d-1}}
\int_{\alpha}^{1}
\mathrm{VaR}_u(X_i)\,
\frac{\theta e^{-\theta u}}{1-e^{-\theta u}}
\left[
\ln\!\left(
\frac{e^{-\theta u}-1}{e^{-\theta \alpha}-1}
\right)
\right]^{d-2}
\, du,
\qquad \alpha \in(0,1),	
\end{equation}
\end{Prop}
\begin{Proof}
The proof follows the same steps as Proposition~\ref{VaRC} by replacing the generator with that of the Frank copula and is therefore omitted.
\end{Proof}
\subsection{Gumbel-Hougaard's family of copulas and the  VaR}
The Gumbel-Hougaard's family of copulas is a statistical model designed to capture the dependence structure between multiple variables, particularly emphasizing their joint behavior in the tails or during extreme events. This family of copulas is mainly used to describe upper-tail dependence, especially in cases of extreme gains or losses of assets.

Its definition and corresponding generator, according to  Nelsen (2006) and Nadarajah et al. (2017), are  as follows. 
\begin{Def}
For all  $(u_{1},\cdots,u_{d})\in[0,1]^{d}$ et $\theta>1$, the  Gumbel-Hougaard's family of copulas is defined by
\[ 
C_{\theta}(u_{1},\cdots,u_{d})=exp\biggl\{-\biggl[\sum_{i=1}^{d}(-lnu_{i})^{\theta}\biggr]^{\frac{1}{\theta}}\biggr\},
\]
with associated generator function expressed as	
\begin{equation}
	\phi_{\theta}(t)=(-ln t)^{\theta}, \; \forall t\in (0,1],  \; \theta \in[1,\infty).
\end{equation}	
The corresponding tau of Kendall  is given by  $\tau (\theta)=1-1/\theta,\;\theta \in[1,\infty)$ .
\end{Def}

\begin{Rem}
The Gumbel-Hougaard copula  models  positive  dependence only. It shows asymmetric tail behaviour and upper tail dependence, with independence recovered at $\theta=1$ and complete dependence attained as   $\theta \to +\infty$ (Nadarajah et al., 2017).  
\end{Rem}

The multivariate   Value-at-Risk is given by the following proposition~\ref{VaRGumb}.
\begin{Prop}~\label{VaRGumb}
Let $\textbf{X} = (X_1, \dots, X_d)$ be a random vector with joint distribution F, satisfying regularity conditions and associated with a Gumbel-Hougaard's family of copulas whose generator is given by $\phi_{\theta}(t) = [-ln (t)]^{\theta}, \; \forall t\in (0,1]$ for $ \theta \geq 1$. 
Then, $\forall i \in \{1,\cdots,d \}$
	\begin{equation}
	\text{VaR}_{\alpha}^{i}(\textbf{X})=\frac{\theta(d-1)}{[-ln(\alpha)]^{\theta(d-1)}}\int_{0}^{-ln\alpha}VaR_{\exp(-u)}(X_{i})\times u^{\theta-1}[(-ln\alpha)^{\theta}-u^{\theta}]^{d-2}du,\; \alpha \in (0,1)
	\end{equation}
with dependence parameter $\theta \in [1,\infty)$.
\end{Prop}

The proof of Proposition~\ref{VaRGumb} is provided in~\ref{supplApp}.
\subsection{Joe's family of copulas and the  VaR}
The Joe's family of copulas is particularly characterized by its ability to capture strong upper tail dependence while exhibiting no lower tail dependence. This asymmetric dependence structure makes the Joe copula especially suitable for modeling joint extreme events occurring in the upper tails of distributions, such as simultaneous large losses or extreme market movements. This family of copulas models  positive dependence only (Nadarajah et al., 2017). A recent application of copulas due to Joe is portfolio risk analysis with Asian equity markets (Nadarajah et al., 2017)

\begin{Def}
For all  $(u_{1},\dots,u_{d}) \in [0,1]^{d}$ and $\theta \geq 1$, the $d$ dimensional Joe's family of copulas is  expressed as (Nelsen, 2006; Niekiema, 2024), 
	\[ 
	C_{\theta}(u_{1},\dots,u_{d})=1-\biggl[\prod_{i=1}^{d}(1-u_{i})^{\theta}\biggr]^{\frac{1}{\theta}},
	\]
with generator function (Nelsen, 2006),
\begin{equation}  
\phi_{\theta}(t)=-ln[1-(1-t)^{\theta}],\;\forall t\in [0,1],\;\theta \in [1, \infty).
\end{equation} 
	
The related tau of Kendall is given by $
\tau (\theta)=1+(4/\theta) \int_{0}^{1} \{[1-(1-t)^{\theta}]ln[1-(1-t)^{\theta}] \}/(1-t)^{\theta -1}dt,\;\theta \in [1, \infty)$ (Nelsen, 2006; Thilini, 2022).
\end{Def}

\begin{Rem}
This copula captures strong positive dependence with pronounced upper tail dependence. Independence is obtained at $\theta= 1$, while complete dependence is achieved as $\theta \to \infty$.
\end{Rem}

\begin{Prop}~\label{VaRJoe}
Let $\textbf{X} = (X_1, \dots, X_d)$ be a random vector with joint distribution F, satisfying regularity conditions and associated with Joe's family of copulas whose generator is given by  $\phi_{\theta}(t)=-ln[1-(1-t)^{\theta}],\;\forall t\in [0,1],\; \mathrm{for}\;\theta \in [1, \infty)$.
Then, $\forall i \in \{1,\cdots,d \}$
\begin{equation}
\mathrm{VaR}_{\alpha}^{i}(\textbf{X})=\frac{\theta(d-1)}{\biggl\{-ln[1-(1-\alpha)^{\theta}]\biggr\}^{d-1}}\int_{0}^{1-\alpha}VaR_{1-u}(X_i)\frac{u^{\theta-1}}{1-u^{\theta}}
		\biggl\{ln\biggl[\frac{1-u^{\theta}}{1-(1-\alpha)^{\theta}}\biggr]\biggr \}^{d-2}du,\;\forall  \alpha \in (0,1),	
\end{equation}
with $\theta \in [1, \infty).$ 
\end{Prop}

The proof of Proposition~\ref{VaRJoe} can be found in~\ref{supplApp}.
\subsection{Ali-Mikhail-Haq's family of copulas and the  VaR}
The Ali–Mikhail–Haq (AMH) copula is commonly used to model weak to moderate dependence structures. It exhibits limited dependence and does not present tail dependence (Nelsen, 2006; Joe, 1997). Recent applications include parameter estimation in coherent reliability systems  (Nadarajah et al., 2017).

Although formal multivariate extensions of the AMH copula have been proposed, the copula does not admit a genuine Archimedean extension in arbitrary dimension, since its generator is not completely monotone on $[0,1]$. Herein, we therefore restrict our attention to the bivariate case which is
proved to be a genuine Archimedean copula (Nelsen, 2006).

\begin{Def}
	For all  $(u_{1},u_{2})\in[0,1]^{2}$ and $\theta  \in [-1, 1]$, the $2$-variate  AMH's family of copulas is expressed as 
	\[
	C_{\theta}(u_{1},u_{2})= \frac{u_{1}u_{2}}{ 1-\theta (1-u_{1}) (1-u_{2}) }.
	\]
The linked  generator function is expressed as (Nelsen, 2006) 
\begin{equation} 
		\phi_{\theta}(t)=ln\biggl [\frac{1-\theta(1-t)}{t} \biggl ],\;\forall t\in (0,1],\;\theta \in [-1, 1],
\end{equation}
while  the tau of Kendall is given  
$ \tau (\theta)=1-(2/3) [\theta + (1-\theta)^{2}ln(1-\theta)]/\theta^{2},\;\theta \in [-1, 1]$ (Nelsen, 2006; Thilini, 2022).
\end{Def}

\begin{Rem}
The AMH  allows for weak to moderate positive and negative dependence. It is symmetric and does not exhibit tail dependence. Independence is obtained as $\theta \to 0$, while the copula does not attain complete dependence for any finite value of $\theta$.
\end{Rem}
\begin{Prop}~\label{VaRAMH}
Let $\textbf{X} = (X_1, X_2)$ be a bivariate random vector with joint distribution F, satisfying regularity conditions whose generator is given by  $\phi_{\theta}(t)=ln\{[1-\theta(1-t)]/t \},\;\forall t\in (0,1]$. Then,  $\forall i \in \{1,2\}$
\begin{equation}
\mathrm{VaR}_{\alpha}^{i}(\textbf{X})=\frac{1-\theta}{ln\biggr[\frac{1-\theta(1-\alpha)}{ \alpha}\biggr] }  \int_{\alpha}^{1}VaR_{u}(X_{i})\frac{1}{u[1-\theta(1-u)]}du,\; \alpha\in (0,1).
\end{equation}
\end{Prop}
\begin{Proof}
The proof is obtained by adapting the arguments of Proposition~\ref{VaRC} to the Ali–Mikhail–Haq copula through its generator and is omitted for brevity.
\end{Proof}

\section{Illustrative Application: Simulation study}\label{IllusVarApp}

To assess the finite-sample performance of the proposed multivariate Value-at-Risk (VaR) formulas, a Monte Carlo simulation study was conducted using the \texttt{copula} package (Hofert et al., 2025) in the \textsf{R} environment (\textsf{R} Core Team., 2024). For each Monte Carlo replication, a random sample of size $n$ was generated from a given Archimedean copula with uniform margins on $[0,1]$, and the multivariate VaR estimator was computed as a conditional expectation on a neighborhood of the copula level set at confidence level $\alpha$.

The estimator is defined by
\begin{equation}
\widehat{\mathrm{VaR}}_{n,\alpha}
=
\frac{1}{|S_\alpha|}
\sum_{k \in S_\alpha} \mathbf{X}^{(k)},
\qquad
S_\alpha =
\left\{
k \in \{1,\dots,n\} :
\left| C(\mathbf{U}^{(k)}) - \alpha \right| \le h
\right\},
\label{eq:VaR_estimator_MC}
\end{equation}
where $h>0$ is a tolerance parameter. This parameter controls the width of the neighborhood around the copula level set $\partial L^C_\alpha$ used to approximate the conditional expectation defining the multivariate VaR. The Monte Carlo VaR estimate is obtained by averaging over $M=1000$ independent replications.

The simulation study focuses on a trivariate random vector $\mathbf{X}=(X_1,X_2,X_3)$ and considers Clayton, Frank, Gumbel-Hougaard, Joe copulas.  In each case, the dependence parameter is calibrated so that Kendall’s tau equals $\tau=0.5$, ensuring a comparable overall dependence strength across copula families. The confidence level and tolerance parameter are fixed at $\alpha=0.05$ and $h=10^{-4}$, respectively, and sample sizes range from $n=5\times10^4$ to $n=10^6$.  The choice $h = 10^{-4}$ reflects a bias-variance trade-off adapted to the sample sizes considered ($n \in \{5\times 10^4,1 \times 10^5, 5\times 10^5,1\times 10^6\}$): it is small enough to ensure a local approximation of the level set, while remaining large enough to include a sufficient number of observations for numerical stability. Empirically, this value yields stable estimates with negligible bias, as confirmed by the convergence results reported in Table~\ref{tab:MC_All_Copulas_VaR}.

Theoretical VaR values are computed from the closed-form expressions derived in Section~\ref{ArchCp}, with numerical integration performed using standard routines in \textsf{R}. Table~\ref{tab:MC_All_Copulas_VaR} documents the finite-sample convergence of the proposed multivariate VaR estimator. The mean estimated VaR across Monte Carlo replications, together with the standard deviation, absolute bias relative to the theoretical VaR, and the root mean squared error (RMSE) are recorded. For all copula families considered, the results clearly indicate convergence of the Monte Carlo estimator toward the theoretical VaR as the sample size increases. In particular, the standard deviation and RMSE decrease monotonically with the sample size $n$, while the bias remains small even for moderate sample sizes and becomes negligible for large samples. These findings confirm the numerical stability and consistency of the estimator. Moreover, despite being calibrated to the same Kendall’s tau, the copulas yield markedly different VaR levels, reflecting the impact of tail dependence on joint risk. The higher VaR values obtained under the Joe and Gumbel–Hougaard copulas illustrate the amplifying effect of strong upper-tail dependence, whereas the Clayton and Frank copulas produce comparatively lower risk levels. Overall, Table~\ref{tab:MC_All_Copulas_VaR} highlights both the accuracy of the proposed estimator and the critical role of dependence structure in multivariate risk assessment.

Unlike fully simulation-based approaches, which rely on repeated numerical approximations of the copula level set, the proposed methodology benefits from closed-form VaR expressions. This significantly reduces computational complexity and allows Monte Carlo simulations to be used solely for validation purposes, resulting in substantial efficiency gains.

\begin{table}[H]
\centering
\small
\begin{tabular}{c c c c c c c}
\hline
Sample size $n$ & Copula & Mean $\widehat{\mathrm{VaR}}_{\alpha}^{\text{mc}}$ & Std. Dev. & Bias & RMSE & $\mathrm{VaR}_{\alpha}^{\mathrm{theo}}$ \\
\hline
$5\times 10^{4}$  & Clayton & 0.124092 & 0.000638 & 0.000132 & 0.000651 & 0.123961 \\
$1\times 10^{5}$  &  & 0.124123 & 0.000378 & 0.000162 & 0.000411 &  \\
$5\times 10^{5}$  &  & 0.124009 & 0.000232 & 0.000048 & 0.000237 &  \\
$1\times 10^{6}$  &  & 0.123965 & 0.000138 & 0.000004 & 0.000138 &  \\
\hline
$5\times 10^{4}$  & Frank   & 0.237850 & 0.001668 & 0.000032 & 0.001668 &0.237818  \\
$1\times 10^{5}$  &   & 0.237998 & 0.000944 & 0.000180 & 0.000961 &  \\
$5\times 10^{5}$  &    & 0.237865 & 0.000818 & 0.000047 & 0.000819 &  \\
$1\times 10^{6}$  &   & 0.237830 & 0.000314 & 0.000012 & 0.000314 &  \\
\hline
$5\times 10^{4}$  & Gumbel-Hougaard  & 0.252042 & 0.001844 & 0.000213 & 0.001857 & 0.251829 \\
$1\times 10^{5}$  &   & 0.252079 & 0.000995 & 0.000251 & 0.001026 &  \\
$5\times 10^{5}$  &   & 0.251779 & 0.000218 & 0.000050 & 0.000223 &  \\
$1\times 10^{6}$  &  & 0.251855 & 0.000158 & 0.000026 & 0.000160 &  \\
\hline
$5\times 10^{4}$  & Joe  &0.317722  & 0.000228 &0.000369& 0.000434  & 0.317353 \\ 
$1\times 10^{5}$  &   & 0.317651 & 0.000547 & 0.000298 & 0.000623 &  \\
$5\times 10^{5}$  &   & 0.317323 & 0.000201 & 3e-05 & 0.000203 &  \\
$1\times 10^{6}$  &  & 0.317416 & 0.000397 & 6.4e-05 & 0.000402 &  \\
\hline
\end{tabular}
\renewcommand{\arraystretch}{1.1}
\setlength{\tabcolsep}{6pt}
\caption{Comparison of convergence of the copula-based multivariate VaR across Clayton, Frank, and Gumbel-Hougaard, Joe  copulas with parameters $\theta=2$, $\theta=5.74$, $\theta=2$ and 
$\theta=2.4$, respectively by setting $\tau=0.5$,  for different sample sizes with $\alpha=0.05$.}
\label{tab:MC_All_Copulas_VaR}
\end{table}

%While the simulation study confirms the theoretical results under controlled settings, the next section illustrates the proposed methodology on real financial data.

\section{Concluding remarks}\label{ConcRem}

This paper proposes explicit analytical expressions for the multivariate lower-tail Value-at-Risk within the class of Archimedean copulas, offering a practical and computationally efficient alternative to purely simulation-based risk assessment methods. By exploiting the generator representation and Kendall’s distribution, closed-form formulas are derived for several widely used copula families, enabling a clear interpretation of how dependence structures affect multivariate risk levels.

Monte Carlo simulations are used to illustrate the practical performance of the proposed approach. The numerical results confirm that the analytical expressions provide accurate and stable VaR estimates in finite samples. Furthermore, the simulations show that copulas calibrated to the same Kendall’s tau may give substantially different VaR values, emphasizing the importance of selecting dependence models that adequately capture tail behavior in applied risk management settings.

From a computational perspective, the availability of closed-form expressions significantly reduces numerical complexity. In practice, this allows Monte Carlo simulations to be employed primarily for validation rather than estimation, resulting in notable gains in efficiency and scalability, especially for large sample sizes.

Beyond their methodological interest, the results of this paper are directly relevant for systemic risk analysis and regulatory risk management frameworks such as Basel~III/IV. In these settings, capital requirements and stress tests are largely driven by tail risk measures computed at high confidence levels, where dependence and joint extreme events play a critical role. The proposed copula-based multivariate VaR formulas provide transparent tools to assess how different dependence structures affect aggregate risk, by  such means helping to identify sources of systemic vulnerability that may be underestimated under simplified or Gaussian dependence assumptions. In particular, the numerical results show that copulas calibrated to the same Kendall’s tau can lead to markedly different VaR levels due to their tail dependence properties, a feature that is highly relevant for stress-testing exercises and scenario analysis.  The proposed framework may consequently complement existing regulatory tools by facilitating sensitivity analysis, model comparison, and robustness checks in multivariate risk assessment, through offering tractable and computationally efficient expressions.

Several directions for further applied research emerge from this work. The methodology can be extended to other multivariate risk measures, such as multivariate Expected Shortfall, which is of direct relevance for regulatory and managerial applications. Further work may also consider more general marginal distributions, higher-dimensional portfolios, and time-varying dependence structures to better reflect real-world financial and insurance data. These extensions would enhance the applicability of the proposed framework in practical multivariate risk management problems.

\section*{Acknowledgements}

The authors are grateful for useful discussions that contributed to the development of this work.

\newpage
\appendix
\section{Proofs of Proposition~\ref{VaRGumb} and Proposition~\ref{VaRJoe} }~\label{supplApp}% Sera %numérotéA 

\begin{Proof}
For $\alpha\in (0,1)$ and $\theta \in [1,\infty)$, one has $\forall t\in (0,1],\;
\phi_{\theta}(t)=(-ln t)^{\theta}\; \mathrm{and} \; \phi^{\prime}(t)=-\theta (-ln t)^{\theta -1}/t$. Then, $\beta_{d}(u,\alpha)=-\phi^{\prime}(u)[\phi(\alpha)-\phi(u)]^{d-2}=\theta(-ln u)^{\theta -1}[(-ln\alpha)^{\theta}- (-ln u)^{\theta}]^{d-2}/u.$
Thus,
\[
\mathrm{VaR}_{\alpha}^{i}(\textbf{X})=\frac{d-1}{\phi(\alpha)^{d-1}}\int_{\alpha}^{1}VaR_u
		(X_i)\frac{\theta}{u}(-ln u)^{\theta -1}[(-ln \alpha)^{\theta}- (-ln u)^{\theta}]^{d-2}du	
\]		
Setting $t=-lnu$ implies  $u=\exp(-t)\; \mathrm{and} \; du=-\exp(-t)dt$. Finally,
\begin{align*}
\mathrm{VaR}_{\alpha}^{i}(\textbf{X})&=\frac{\theta (d-1)}{\phi(\alpha)^{d-1}}\int_{-ln \alpha}^{0} -VaR_{\exp(-u)}(X_i) u^{\theta -1}[(-ln \alpha)^{\theta}- (-ln u)^{\theta}]^{d-2}du \\
		& =	\frac{\theta (d-1)}{\phi(\alpha)^{d-1}}\int_{0}^{-ln \alpha}VaR_{\exp(-u)}(X_i) u^{\theta -1}
		[(-ln \alpha)^{\theta}- (-ln u)^{\theta}]^{d-2}du,	
\end{align*}
providing the desired result.
\end{Proof}

\begin{Proof}
For $\alpha\in (0,1)$ and $\theta \in [1, \infty)$,  $\forall t\in [0,1],\;\phi_{\theta}(t)=-ln[1-(1-t)^{\theta}]$ and  $\phi_{\theta}^{\prime}(t)=-[\theta(1-t)^{\theta-1}]/[1-(1-t)^{\theta}].$ 
Then,
	\begin{align*}
		\beta_{d}(u,\alpha)&=-\phi^{\prime}(u)\biggl[\phi(\alpha)-\phi(u)\biggr]^{d-2}\\
		&=\frac{\theta (1-u)^{\theta -1}}{1-(1-u)^{\theta}}\biggl\{ln\biggl[\frac{1-(1-u)^{\theta}}{1-(1-\alpha)}\biggr]\biggr\}^{d-2}
	\end{align*}
It comes that
\begin{align*}
\mathrm{VaR}_{\alpha}^{i}(X_{i})=\frac{d-1}{\phi(\alpha)^{d-1}}\int_{\alpha}^{1}VaR_u (X_i)\frac{\theta (1-u)^{\theta -1}}{1-(1-u)^{\theta}}\biggl\{ln\biggl[\frac{1-(1-u)^{\theta}}{1-(1-\alpha)}\biggr]\biggr\}^{d-2}du.
\end{align*}
Putting  $t=1-u$ leads to
\begin{align*}
		\mathrm{VaR}_{\alpha}^{i}(X)&=\frac{d-1}{\phi(\alpha)^{d-1}}\int_{1-\alpha}^{0}VaR_{1-t}(X_i)\frac{\theta t^{\theta -1}}{1-t^{\theta}}\biggl\{ln\biggl[\frac{1-t^{\theta}}{1-(1-\alpha)^{\theta}}\biggr]\biggr\}^{d-2}(-dt)\\
		&=\frac{\theta (d-1)}{\phi(\alpha)^{d-1}}\int_{0}^{1-\alpha}VaR_{1-t}(X_i)\frac{\theta t^{\theta -1}}{1-t^{\theta}}\biggl\{ln\biggl[\frac{1-t^{\theta}}{1-(1-\alpha)^{\theta}}\biggr]\biggr\}^{d-2}dt,
\end{align*}	
giving the announced result.	
\end{Proof}

%\end{appendices}

\begin{thebibliography}{99} 	
\bibitem[Barbe et al.(1996)]{BGGR99} Barbe, P., Genest, C., Ghoudi, K. and Rémillard, B. (1996). On Kendall’s process. \textit{Journal of Multivariate Analysis}, 58(2), 197-229
	
\bibitem[Belzunce et al.(2007)]{BCOS07} Belzunce, F., Castaño, A., Olvera-Cervantes, A., and Suárez-Llorens, A. (2007). Quantile curves and dependence structure for bivariate distributions. \textit{Computational Statistics and Data Analysis}, 51(10), 5112-5129


\bibitem{CB13} Cousin, A., and Di Bernardino, E. (2013). On multivariate extension of Value-at-Risk. \textit{Journal of Multivariate Analysis}, \textbf{119}, 32-46.
			
\bibitem{DB11} Di Bernardino, E. (2011). \textit{Modélisation de la dépendance et mesures de risque multidimensionnelles} [Thèse de doctorat, Université Claude Bernard Lyon 1]. HAL Archives. \url{https://hal.science/tel-00638382}

\bibitem{DQ07} Durante, F., and Quesada-Molina, C. (2007). A generalization of the Archimedean class of bivariate copulas. \textit{Annals of the Institute of Statistical Mathematics}, 59(3), 487–498.

\bibitem{EP06} Embrechts, P., and Puccetti, G. (2006). Bounds for functions of multivariate risks. \textit{Journal of Multivariate Analysis}, 97(2), 526-547

\bibitem{GR01} Genest, C., and Rivest, L.-P. (2001). On the multivariate probability integral transformation. \textit{Statistics and Probability Letters}, 53(4), 391-399


\bibitem{GH19} Guégan, D., and Hassani, B. K. (2019). Risk measurement: From quantitative measures to management decisions. Springer International Publishing

\bibitem{HKMY25} Hofert, M., Kojadinovic, I., M{\"a}chler, M., and Yan, J. (2025).
\textit{copula: Multivariate dependence with copulas} (R package version 1.1-6).
\url{https://CRAN.R-project.org/package=copula}



\bibitem{HKM18} Hofert, M., Kojadinovic, I., and M{\"a}chler, M. (2018).
Elements of copula modeling with R. Cham, Switzerland: Springer

\bibitem{HMS18} Hoffmann, H., Meyer-Brandis, T. and Svindland, G. (2018). Strongly consistent multivariate conditional risk measures. \textit{Mathematics Financial Economics}, 12, 413-444. 


\bibitem{HW14} H{\"u}rlimann, W. (2014). On some properties of two vector-valued VaR and CTE multivariate risk measures for Archimedean copulas. ASTIN Bulletin: \textit{The Journal of the IAA}, 44(3), 613-633

\bibitem{JH97} Joe, H. (1997). \textit{Multivariate models and dependence concepts} (Vol. 73). Monographs on Statistics and Applied Probability. Chapman and Hall, London


\bibitem{LP19} Lux, T., and Papapantoleon, A. (2019). Model-free bounds on Value-at-Risk using extreme value information and statistical distances. \textit{Insurance: Mathematics and Economics}, 86, 73-83.


\bibitem{NAC17} Nadarajah, S., Afuecheta, E., and Chan, S. (2017). A compendium of copulas. \textit{Statistica}, 77(4), 279-328

	
\bibitem{NS09} Nappo, G., and Spizzichino, F. (2009). Kendall distributions and level sets in bivariate exchangeable survival models. \textit{Information Sciences}, 179(18), 2878-2890.
	
\bibitem{NR06} Nelsen, R. B. (2006). An introduction to copulas (2nd ed.). Springer. (Springer Series in Statistics).

\bibitem{NW25} Nikiéma, W. B. (2025).
\textit{Copules et modèles des extrêmes en finance stochastique}
[Thèse de doctorat, Université Joseph Ki-Zerbo, Ouagadougou, Burkina Faso].


\bibitem{QSWY19} Qian, L., Shen, Y., Wang, W., and Yang, Z. (2019). Valuation of risk-based premium of DB pension plan with terminations. \textit{Insurance: Mathematics and Economics}, \textbf{86}, 51-63

\bibitem{R24}  R Core Team. (2024). R: \textit{A language and environment for statistical computing}. R Foundation for Statistical Computing, Vienna, Austria.
	
\bibitem{SM59} Sklar, M. (1959). Fonctions de répartition à $n$ dimensions et leurs marges. \textit{Publications de l’Institut de Statistique de l’Université de Paris}, \textbf{8}, 229–231.

\bibitem{TL93} Tibiletti, L. (1993). On a new notion of multidimensional quantile. \textit{Metron- International Journal of Statistics}, 51(1), 77-83
	
\bibitem{TD22} Thilini, D. K. (2022).
\textit{Modern statistical methodologies in mortality forecasting and insurance applications}
[Doctoral dissertation, Macquarie University, Sydney, Australia].
	
\end{thebibliography}
\end{document}